\documentclass[twocolumn,superscriptaddress,showpacs,pra]{revtex4}
\usepackage{epsfig}
\usepackage{delarray}
\usepackage{amsmath, amssymb}
\usepackage{bm}
\usepackage{graphicx}
\usepackage{dcolumn}
\usepackage{bm}

\begin{document}
\title{Parameter Estimation with Dzyaloshinskii-Moriya Interaction \\ under External Magnetic Fields}

\author{Fatih Ozaydin}
\email{mansursah@gmail.com}
\affiliation{Department of Information Technologies, Isik University, Istanbul, Turkey}

\author{Azmi Ali Altintas}
\affiliation{Department of Electrical Engineering, Faculty of Engineering and Architecture, Okan University, Istanbul, Turkey}

\begin{abstract}
We study the effects of external magnetic fields on the precision of parameter estimation with thermal entanglement of two spins in XX model, in the presence of Dzyaloshinskii-Moriya (DM) interaction.
Calculating the quantum Fisher information, we show that homogeneous magnetic field $B$, inhomogeneous magnetic field $b$ or DM interaction $D$ increases the precision of parameter estimation, overwhelming the destructive effects of thermalization.
We also show that for the model in consideration, the effects of $b$ and $D$ are the same.
However, the existence of both $b$ and $B$ (or both $D$ and $B$) decreases the precision.
We find that in order to increase the precision in parameter estimation tasks, applying $b$ in the ferromagnetic case and $B$ in the antiferromagnetic case should be preferred.
\end{abstract}

\maketitle

\section{Introduction}

Dzyaloshinskii-Moriya interaction \cite{Dz1958,MoriyaPRL} plays an important role in various fields from spin Hall effect \cite{Noel2014I3EMag}, skyrmions \cite{Yoo2014I3EMag,Jalil2014JAP,OuYang2014}, magnetic domain walls \cite{Janutka2015} and thin films \cite{Nembach2015,Levente2014} to spin glasses \cite{Lyakhimets93}.
Furthermore, with the discovery that Dzyaloshinskii-Moriya (DM) interaction excites the entanglement of a two qubit XYZ spin chain \cite{Zhang2007}, an intense effort has been devoted to explore the entanglement dynamics of Heisenberg spin models of two and three level systems with DM interaction, with and without external magnetic fields \cite{Must2008OptComm,Jaferi2008PRB,Langari2009PRA,Jafari2014PRA,Yea2014PhysA,Sharma2014,Sharma2015}.
These works focused on how DM interaction overwhelms the disentangling effects of thermalization and even external magnetic fields.

Besides secret key distribution and speed-up in algorithms \cite{Nielsen}, a key application area that quantum mechanics can provide advantages is metrology, where the precision of estimating the parameter $\phi$ of an $N$ particle state $\rho(\phi)$ is limited by the quantum Cram\'{e}r-Rao bound $\Delta \phi_{QCB} \equiv 1 / \sqrt{ N_m F_q}$ with $N_m$ the number of measurements and $F_q$ the quantum Fisher information of the state.
In particular, for a single measurement i.e. $N_m=1$ and a state of $N$ particles, $F_q \leq N$ for separable states, achieving the shot noise limit (SNL) $\Delta \phi_{SNL} = 1 / \sqrt{N}$.
However entangled states can achieve  $F_q \leq N^2$, implying $\Delta \phi_{HL} \equiv 1 / N$ which is called the Heisenberg Limit (HL) \cite{PezzeSmerzi2009PRL}.
Playing such a central role in metrology, there has been an increasing attraction on quantum Fisher information, usually maximized over the directions $x$, $y$ and $z$, and averaged per particle \cite{Ma2011PRA}, which we denote here as $QFI$.
In this setting, SNL reads $QFI=1$ and HL reads $QFI = N$.

For a state $\rho$, surpassing the shot noise limit, i.e. $QFI(\rho)>1$ witnesses entanglement and such states are called \textit{useful} states but the converse is not always true: An entangled state may not surpass SNL \cite{Hyllus2012PRA}.
Two states of same entanglement can have different values of QFI, and unlike entanglement measures which are monotonic by definition, QFI of a state can increase under local unitary operations \cite{Erol2014SciRep,Hyllus2010PRA}.
What is more, there are cases that as the entanglement of a state decreases, its QFI increases \cite{Ma2011PRA}.
Being such deeply related but not directly proportional to entanglement, QFI has been studied for classes of multiparticle entangled states \cite{Toth2015NJP,Ma2011PRA,Ozaydin2014PLA,Ozaydin2014IJTP,Jaksch2009PRA,Ozaydin2015IJTP,Ozaydin2015ActaQFI}
 spin chains \cite{Wang2013JPA,LiuDuLiu2016}, Lipkin-Meshkov-Glick model \cite{LMG} and an open dissipative system with reset mechanism \cite{AoP}.
Recently, we have studied the quantum Fisher information of the thermal entanglement of the XXX model and shown that DM interaction can overwhelm the thermalization effects \cite{Ozaydin2015SciRep1}.
When it comes to the effect of external magnetic fields on the entanglement of Heisenberg spin chains, it has been found that this effect is usually destructive \cite{Must2008OptComm,Zhou2008PhysScr,Cao2009OptComm}, with a striking counterexample that magnetic fields excite the entanglement of the antiferromagnetic Heisenberg spin chain \cite{VedralPRL2001}.
Therefore it becomes interesting to consider external magnetic fields in quantum metrology especially in the presence of DM interaction.

In this work, we focus on this problem.
We study the effects of external magnetic fields on the precision of parameter estimation with XX model in the presence of DM interaction, by calculating the quantum Fisher information.
We show that -on the contrary to entanglement in general-, external magnetic fields can excite the quantum Fisher information of the Heisenberg model, i.e. increasing the precision of parameter estimation.
In particular, we show that in the model we consider, either $B$, $b$ or $D$ alone increases the QFI, the effects of $b$ and $D$ are the same and the existence of $b$ and $B$ (or $D$ and $B$) decreases the QFI.
We also show that, $b$ in the ferromagnetic case and $B$ in the antiferromagnetic case should be preferred to increase the QFI. The Hamiltonian of the model we study can be described by ${1 \over 2} \sum_{i=1}^{N} \{J[\sigma_{x}^{i}\sigma_{x}^{i+1} + \sigma_{y}^{i}\sigma_{y}^{i+1}
 + \overrightarrow{D} \cdot (\overrightarrow{\sigma}^{i} \times \overrightarrow{\sigma}^{i+1})]
  + (B+b)\sigma_{z}^{i}    + \ (B-b)\sigma_{z}^{i+1}  \},$ where $J$ is the coupling constant, $N$ is the number of particles, $\sigma_{\alpha}^{i}$ are the Pauli operators $\alpha \in \{x,y,z\}$ applied to $i$'th particle, $B$ and $b$ are the homogeneous and inhomogeneous external magnetic fields respectively and $\overrightarrow{D}$ the strength of DM interaction. For a two spin model with $\overrightarrow{D}= D \overrightarrow{z}$ for simplicity, the effective Hamiltonian is reduced to
\begin{eqnarray}
H_{DM}  =  {1 \over 2} \{J[\sigma_{x}^{1}\sigma_{x}^{2} + \sigma_{y}^{1}\sigma_{y}^{2} +
 D(\sigma_{x}^{1}\sigma_{y}^{2} - \sigma_{y}^{1}\sigma_{x}^{2})]
  \\ + (B+b)\sigma_{z}^{1}    + \ (B-b)\sigma_{z}^{2}  \}. & \nonumber
\end{eqnarray}

\noindent The eigenvalues and the associated eigenvectors of the Hamiltonian $H_{DM}$ are found as
\begin{center}
$\{-B, (0,0,0,1)^T\}$,\\
$\{B,(1,0,0,0)^T\}$,\\
$\{-\gamma, {1 \over N_1}(0,- { i (-b + \gamma) \over J (i + D)},1,0)^T\}$, and \\
$\{\gamma, {1 \over N_2}(0, { i (b + \gamma) \over J (i + D)},1,0)^T\}$,\\
\end{center}
where $N_1 \! = \! \sqrt{ 1 + \left|  { \gamma -b \over J (i + D)} \right| ^2 }$,
$N_2 \! = \! \sqrt{ 1 + \left|  { \gamma + b \over J (i + D)} \right| ^2 }$ and
$\gamma=\sqrt{ b^2 + J^2(1+ D^2)}$.\\

\noindent The density matrix of the thermal entangled state is found as $\rho = e^{- H_{DM} / kT } / Tr(e^{- H_{DM} / kT })$, where $k$ is the Boltzman constant and $T$ is the temperature.
Thus we obtain the density matrix of the system as \\

\noindent
$\left(
  \begin{array}{cccc}
    {1 \over 1 + e^{ 2B \over T} + 2 e^{ B \over T} \gamma_c }      & 0 & 0 & 0 \\
    0 & {\gamma_c - { b \gamma_s \over \gamma } \over 2(\cosh{B \over T} + \gamma_c) }    & { i(i-D) J \gamma_s  \over 2\gamma (\cosh{B \over T} + \gamma_c ) }             & 0 \\
    0 & { i(i+D) J \gamma_s  \over 2\gamma (\cosh{B \over T} + \gamma_c ) }               & {\gamma_c + { b \gamma_s \over \gamma } \over 2(\cosh{B \over T} + \gamma_c) }  & 0 \\
    0 & 0 & 0 & { e^{ B \over T} \over 2(\cosh{B \over T} + \gamma_c)  } \\
  \end{array}
\right)$\\

\begin{figure}[t!]
\includegraphics[width=0.5\textwidth]{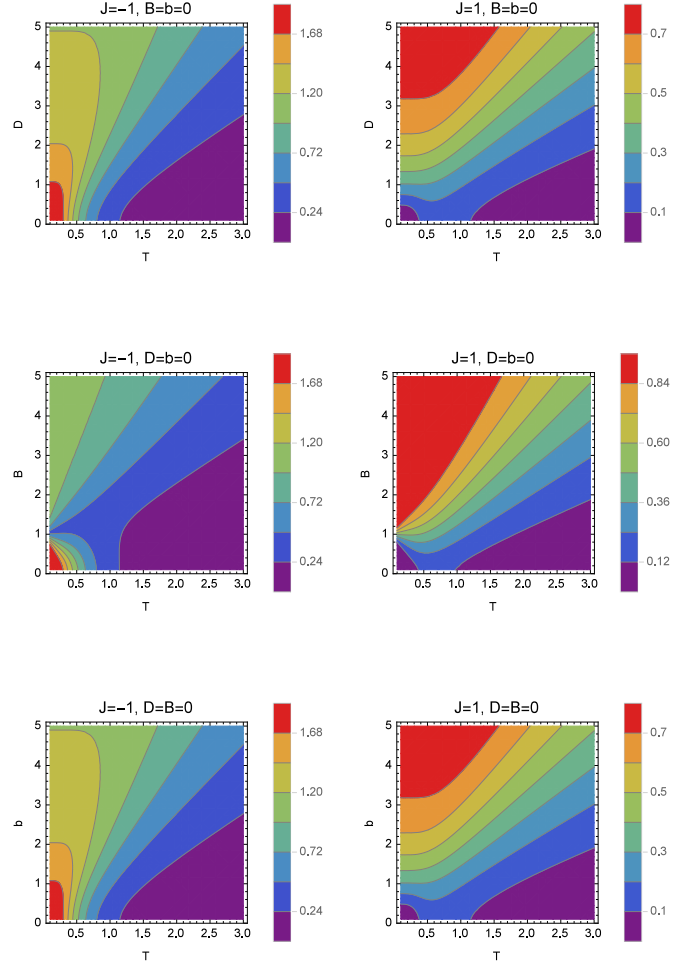}
\caption{Quantum Fisher information per particle (QFI) for ferromagnetic (left column) and antiferromagnetic (right column) Heisenberg chains in the unit of Boltzmann constant $k$ as functions of, (first row) temperature $T$ versus DM interaction $D$  with no external magnetic fields; (second row) temperature $T$ versus homogeneous external magnetic field $B$ with no inhomogeneous external magnetic field and DM interaction; (third row) temperature $T$ versus inhomogeneous external magnetic field $b$ with no homogeneous external magnetic field and DM interaction.}  \label{fig:fig1}
\end{figure}

\noindent with $\gamma_c = \cosh{\gamma \over T}$ and $\gamma_s = \sinh{\gamma \over T}$.
The eigenvalues and the eigenvectors, $\{p_i, | \psi_i\rangle\}$ of the density matrix $\rho$ are used to calculate the QFI of the system.

The parameter to be estimated is usually considered as being acquired by the system $\rho$ via the unitary evolution
$e^{i \phi J_{\vec{n}}} \rho e^{-i \phi J_{\vec{n}}}$, -for example as a phase shift on one arm of a Mach Zender interferometer-
where
$J_{\vec{n}} \equiv \vec{J} \cdot \vec{n} = {1 \over 2} \sum ^N_{\alpha=x,y,z} n_{\alpha}\sigma_{\alpha}^{i}$
with $\sigma^{(i)}$ the Pauli spin operator on $i$th particle.
In our scenario, we consider that the instances of the density matrix $\rho$ with the given values of $J$, $T$, $B$, $b$ and $D$, acquire a parameter $\phi$ and we calculate the QFI of each instance to analyze the effects of the given values on the precision of estimating the parameter $\phi$.
There are a few calculation methods of $F_q$ \cite{Liu2013PRA,Liu2014CTP,Jing2015PRA} and we will use a common one for full-rank density matrices of mixed states \cite{Ma2011PRA}, i.e. $QFI(\rho)= c_{max} / N$ where $c_{max}$ is the largest eigenvalue of the 3 by 3 matrix $C$ constructed using the eigenvalues and eigenvectors of $\rho$ and $J_{\vec{n}}$ as

\begin{equation}C_{kl}\!\! =\!\!\! \sum_{i\neq j}\!\! { (p_i\!\! -\!\! p_j)^2 \over (p_i\!\! +\!\! p_j) } [ \langle \psi_i | J_k | \psi_j \rangle \langle \psi_j | J_l | \psi_i \rangle + \langle \psi_i | J_l | \psi_j \rangle \langle \psi_j | J_k | \psi_i \rangle ],
\end{equation}

\noindent where $k,l \in {x,y,z}$. We find that for the system in consideration, the only non-zero terms of $C$ matrix are $C_{xx}=C_{yy}$, i.e. this system has zero phase sensitivity in $z$ direction and the phase sensitivities in $x$ and $y$ directions are equal to each other, resulting $QFI(\rho)=C_{xx}/N$. We first analyze how each of DM interaction, homogeneous or inhomogeneous magnetic field excites QFI separately, surpassing the destructive effects of thermalization.
We study QFI as a function of each effect with respect to temperature, setting the strength of two other effects to zero for a clear observation.
Due to lengthy terms, we continue with numerical calculations, and as illustrated in Fig.1, show that the existence of any of these three effects excite the QFI of the system in the high temperature region.
An interesting point here is that although resulting in different systems, therefore different corresponding density matrices and different entanglement values, inhomogeneous magnetic field and DM interaction excite the QFI in the same way.
Exhibiting first a sudden increase and then a slow decrease with respect to increasing $B$, $b$ or $D$, the QFI of ferromagnetic chain remains well larger than that of antiferromagnetic chain, surpassing the shot-noise level, in accordance with our recent findings for the XXX model including only DM interaction \cite{Ozaydin2015SciRep1}.

\begin{figure}[t!]
\includegraphics[width=0.5\textwidth]{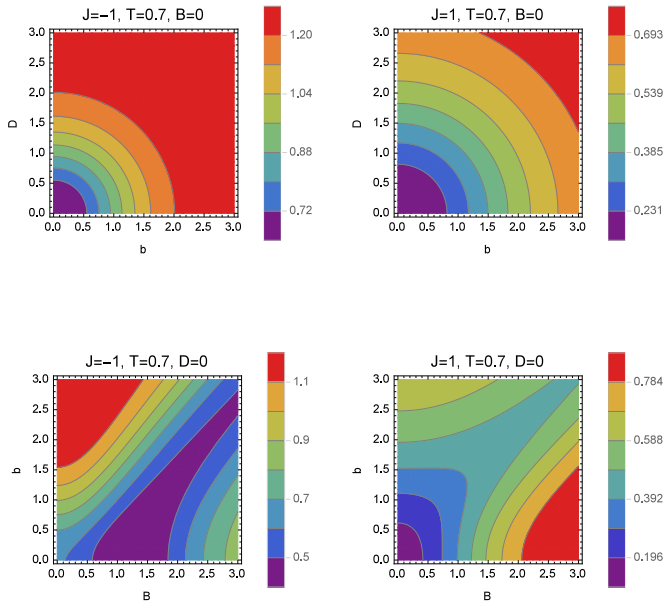}
\caption{Quantum Fisher information per particle (QFI) for ferromagnetic (left column) and antiferromagnetic (right column) Heisenberg chains in the unit of Boltzmann constant $k$ as functions of, (first row) DM interaction $D$ versus $b$  with no external homogeneous magnetic field; (second row) external inhomogeneous magnetic field $b$ versus external homogeneous magnetic field $B$ with no DM interaction. Temperature is fixed at $T=0.7.$} \label{fig:fig2}
\end{figure}

By fixing the temperature at $T=0.7$, we explore the effects of DM interaction and magnetic fields together, and homogeneous and inhomogeneous magnetic fields together in QFI in Fig.2.
We find that, although $D$ and $b$ increase the QFI in the same way, only the ferromagnetic model can surpass the SNL, even at high temperatures.
However, although $b$ or $B$ alone increases the QFI, the existence of both $b$ and $B$ decreases the QFI in an asymmetric way such that in the ferromagnetic case $b$, and in the antiferromagnetic case $B$ should be applied to achieve a higher QFI, implying a higher precision of parameter estimation.
Note that, since the effects of $D$ and $b$ are the same in the model in consideration, we do not plot the contour of QFI with respect to $D$ and $B$.

\section{Conclusion}
We have studied the precision of parameter estimation under external magnetic fields and in the presence of Dzyaloshinskii-Moriya (DM) interaction.
Considering the XX model of two spins, we have found that either homogenous or inhomogeneous magnetic fields, or DM interaction alone increases the precision of parameter estimation.
In the very specific case of our model, the effects of inhomogeneous magnetic field and DM interaction are the same.
However, the precision decreases if both of the magnetic fields, or both homogeneous magnetic field and DM interaction exists.
A key finding presented in this work is that in order to increase the precision of parameter estimation, applying inhomogeneous magnetic field in the ferromagnetic case and homogeneous magnetic field in the antiferromagnetic case should be preferred.
It would be interesting to extend our work to other spin models as well as spin glasses.

\section{Acknowledgement}
The authors greatly acknowledge the financial support of Isik University Scientific Research Fund, Grant No: BAP-14A101.

\end{document}